\documentclass[doublespacing]{elsart}
\usepackage{graphicx}
\usepackage{amssymb}

\begin{document}

\begin{frontmatter}


%\thanks[label2]{}
%\corauth[cor1]{}

\title{Computation of Lyapunov Exponents in
General Relativity}


\author{Xin Wu, }
\author{ Tian-yi Huang }
\ead{tyhuang@nju.edu.cn}

\address{Department of Astronomy, Nanjing University, Nanjing
210093, China }

\begin{abstract}
Lyapunov exponents (LEs) are key indicators of chaos in dynamical
systems. In general relativity the classical definition of LE
meets difficulty because it is not coordinate invariant and
spacetime coordinates lose their physical meaning as in Newtonian
dynamics. We propose a new definition of relativistic LE and give
its algorithm in any coordinate system, which represents the
observed changing law of the space separation between two
neighboring particles (an ``observer'' and a ``neighbor''), and is
truly coordinate invariant in a curved spacetime.
\end{abstract}

\end{frontmatter}

% main text
%\section{}
%\label{}

% \appendix

% \section{}
% \label{}

Chaos is a popular phenomenon in dynamical systems. One of its
main features is the exponential sensitivity on small variations
of initial conditions. The exhibition of chaos in the motion of
Pluto makes it particularly attractive for scientists to
investigate the dynamical behavior of the solar system[1].

In Newtonian mechanics, Lyapunov exponents (LEs), as a key index
for measuring chaos in a dynamical system, can be calculated
numerically with either the variational method[2] or the
two-particle method[3], and sometimes a mixture of the two[4]. The
former is more rigorous but one has to derive the variational
equations of the dynamical equations of the system and integrate
them numerically with the dynamical equations together. The latter
is less cumbersome especially when one wants to compute the
maximum LE only, which is the main index for chaos. We will
discuss the extension of the two particle method to relativistic
models in this letter.

Let ${\bf q}(t)$ and $\dot{{\bf q}}(t)$ be the position and
velocity vector of a dynamical system. As a set of initial
conditions is selected randomly and the corresponding trajectories
are restricted in a compact region in the phase space, the
classical definition of the maximum LE is
\begin{equation}
\lambda_{N}=\lim_{t\rightarrow\infty} \frac{1}{t} \ln
\frac{d(t)}{d(0)},
\end{equation}
where the separation $d$ between two neighboring trajectories is
required to be sufficiently  small so that the deviation vector
$(\Delta {\bf q}, \Delta \dot{\bf q})$ can be regarded as a good
approximation of a tangent vector, and the distance $d(t)$ at time
$t$ is of the form
\begin{equation}
d(t)=\sqrt{\Delta {\bf q}(t) \cdot \Delta {\bf q}(t)+\Delta
\dot{{\bf q}}(t)\cdot \Delta \dot{{\bf q}}(t)}.
\end{equation}
It is the Euclidian distance in the phase space. ${\bf q}$ and
$\dot{\bf q}$ are in different dimension, so one has to carefully
choose their units to assure that both terms in the expression of
$d(t)$ are important. We suggest computing the LE in the
configuration space rather than in the phase space, that is,
\begin{eqnarray}
d'(t) &=&\sqrt{\Delta {\bf q}(t)\cdot\Delta {\bf q}(t)}\\
\lambda'_{N}&=&\lim_{t\rightarrow\infty} \frac{1}{t} \ln
\frac{d'(t)}{d'(0)},
 \end{eqnarray}
We argue that both $\lambda_{N}$ and $\lambda'_{N}$ are effective
in detecting chaotic behavior of orbits since $\Delta \dot{{\bf
q}}(t)=d/dt(\Delta{\bf q}(t))$ and they should have the same
Lyapunov exponents.

In general relativity the above definition and algorithm are
questionable. First, there is no unified time for all the
reference systems. Secondly, the separation of time and space of
the 4-dimensional spacetime is different for different observers.
Furthermore, time and space coordinates usually play book-keeping
only for events and are not necessary with a physical meaning.
Consequently, one would get different values of $\lambda_N$ and
$\lambda'_N$ in different coordinate systems.

Up to now the references for exploring chaos in relativistic
models have been mainly interested in studying the dynamical
behavior of black hole systems[5-8] and the mixmaster
cosmology[9-13]. Most of them adopt the classical definition of
LE. Here we will abandon the variational method in the case of
general relativity for it is rather cumbersome to derive
variational equations. Actually there exist general expressions
for geodesic deviation equations[5] but one has to derive the
complicated curvature tensor. Furthermore, in many cases a
particle does not follow a geodesic. For instance, a spinning
particle in Schwarzschild spacetime doesn't move along a
geodesic[5]. Therefore, we will concentrate on constructing a
revised edition of the two-particle method. The classical
algorithm of LE lacks invariance in general relativity, which
depends on a coordinate gauge and even could bring spurious chaos
in some coordinate systems. The chaotic behavior in the mixmaster
cosmology[9-13] has been debated for decade or so. Using different
time parameterization [14,15], one would get distinct values of
$\lambda_{N}$. Especially, in a logarithmic time, chaos becomes
hidden[12]. Chaos, as an intrinsic nature of a given system,
should not be affected by the choice of a coordinate gauge, and a
chaos indicator, LE, should be defined as invariant under
spacetime transformations. It means that the LE in general
relativity should be expressed as a physical or so called proper
quantity but not a coordinate quantity[16-19].

Now we consider a particle, called ``observer'', moving along an
orbit (not necessary to be a geodesic) in the spacetime with a
metric $ds^{2}=g_{\alpha\beta}dx^{\alpha}dx^{\beta}$. In this
letter Greek letters run from 0 to 3 and Latin letters from 1 to
3. The observer can determine if his motion would be chaotic by
observing whether the {\it proper} distances from his neighbors
are increasing exponentially or not with his {\it proper} time.
Both the distances and time are observables and should not depend
on the choice of a coordinate system. Then we can apply the theory
of observation in general relativity[16] to explore the dynamical
behavior of the observer.

Fig. 1 illustrates the trajectories of the observer and its
neighboring particle called ``neighbor''. The initial condition of
the neighbor is arbitrarily chosen as long as its 4-dimensional
distance from the observer is small enough to assure that the
neighbor is approximately in the tangent space of the observer. In
an arbitrary spacetime coordinate system $x^{\alpha}$, one can
derive the equations of motion of the observer and its neighbor.
Here we have to notice that the independent variable of the
equations has to be the coordinate time $t$ rather than their
proper times $\tau$ because the two particles have different
proper times but one unique independent variable has to be adopted
when integrating numerically their equations of motion together.
At the coordinate time $t$ the observer arrives at the point $O$
with the coordinate $x^{\alpha}$ and 4-velocity $U^{\alpha}$, and
the corresponding proper time of the observer is $\tau$. At the
same coordinate time $t$, its neighbor reaches the point $P$ with
the coordinate $y^{\alpha}$ along another orbit. A displacement
vector $\delta x^{\alpha}=y^{\alpha}-x^{\alpha}$ from $O$ to $P$
should be projected into the local space of the observer. The
space projection operator of the observer is constructed as
$h^{\alpha\beta}= g^{\alpha\beta}+c^{-2} U^{\alpha}U^{\beta}$ ($c$
represents the velocity of light)[16]. $\overrightarrow{OP'}$
represents the projected vector $\delta
x^{\alpha}_{\bot}=h^{\alpha}_{\beta} \delta x^{\beta}$, and its
length $\| \overrightarrow{OP'}\|$ is
\begin{equation}
\Delta L(\tau)=  \sqrt{g_{\alpha \beta} \delta x^{\alpha}_{\bot}
\delta x^{\beta}_{\bot}}  = \sqrt{h_{\alpha \beta} \delta
x^{\alpha}\delta x^{\beta}}.
\end{equation}
Here $g_{\alpha \beta}$ is calculated at $x^{\alpha}$. $\Delta
L(\tau)$ is the proper distance to the neighbor observed by the
observer at his time $\tau$ and it is a scalar. Hence the maximum
LE in general relativity is defined as
\begin{equation}
\lambda_R=\lim_{\tau \rightarrow\infty} \frac{1}{\tau} \ln
\frac{\Delta L(\tau)}{\Delta L(0)}.
\end{equation}
\begin{figure}[htb]
\includegraphics[scale=0.5]{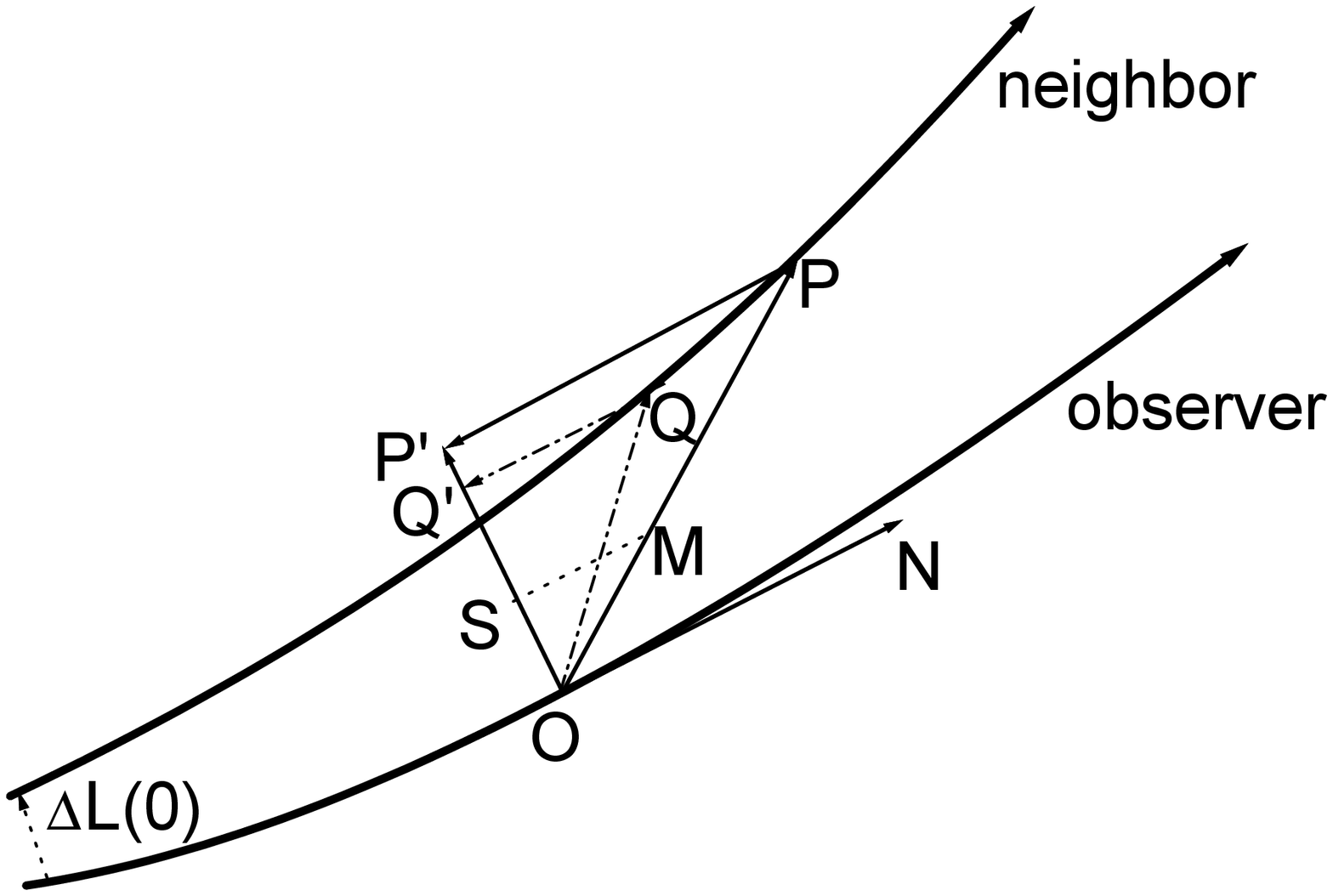}% Here is how to import EPS art
\caption{The trajectories of the observer and its neighbor in
spacetime.}
\label{fig:epsart}
\end{figure}

Let $\Sigma_\tau$ be the 3-dimensional subspace of the tangent
space of the observer at $O$, which is orthogonal to the
4-velocity $U^\alpha$. It is the point $P'$ but not $P$ in
$\Sigma_\tau$ as long as $P'$ is close enough to the observer $O$.
This tells us that $\lambda_R$ is coordinate invariant. The next
is an argument for this point. Let us carry out a time
transformation $t\rightarrow \eta$. For convenience, we express
the projection operation as
$\overrightarrow{OP'}=H\overrightarrow{OP}$, where $H$ is the
space projection operator and $H\overrightarrow{U}_O=0$, where a
subscript is added for $\overrightarrow{U}$ to describe the
4-velocity. Assume that the observer is located at the point $O$
at $\eta$ that corresponds to the old coordinate time $t$, then
the neighbor at $\eta$ would situate at the point $Q$ but not
necessary at $P$. Certainly we have to keep both $P$ and $Q$
inside an $\epsilon$ neighborhood of $O$ (see Fig.1) and
$\epsilon$ is considered as a very small quantity. In this case we
have $\overrightarrow{U}_{P}=\overrightarrow{U}_{O}+O(\epsilon)$.
Furthermore, we have
$\overrightarrow{OQ}=\overrightarrow{OP}-\overrightarrow{QP}$ and
$\overrightarrow{PQ}=\Delta\tau_{P}\overrightarrow{U}_{P}+O(\Delta\tau_P^2)=
k\epsilon \overrightarrow{U}_{P}+O(\epsilon ^2)$, where $k$ is a
constant and the proper time interval of the neighbor between $P$
and $Q$, $\Delta\tau_{P}$, is in the magnitude of $\epsilon$.
Hence, we get the relation
$H\overrightarrow{OQ}=H\overrightarrow{OP}+O(\epsilon^{2})$. This
shows that $\lambda_R$ is invariant with time transformations.

As far as the numerical implementation of the computation of
$\lambda_R$ is concerned, the following notes are worth noticing.
(1) The observer and the neighbor have different proper times, so
a coordinate time $t$ should be adopted as the independent
variable. The equations of the motion of the particles should be
transformed to use $t$ as the independent variable. The variables
to be computed step by step are the space coordinates and
velocities (with respect to $t$) of the observer and its neighbor,
and the proper time $\tau$ and $d\tau/dt$ of the observer. In
total there are 14 variables. (2) As is well known,
renormalization after a certain time interval is essential in this
procedure to keep the distance between the observer and its
neighbor small enough. The renormalization must be proceeded in
the phase space though our LE is calculated in the configuration
space. (3) One important difference during renormalization between
the relativistic and classical cases must be noticed. Let
$\Sigma_t$ be the local 3-space in which all the points have the
same coordinate time $t$. It is evident that the renormalization
should be done in $\Sigma_t$ otherwise the computation will commit
an error. In Fig.1 $P$ is pulled back to $M$ and $
\overrightarrow{OM}=(\Delta L(0)/\Delta L(\tau))
\overrightarrow{OP}$. The next integration step for the neighbor
will start from the point $M$. It is obvious that
$\overrightarrow{OP}$ is located in $\Sigma_t$ because both points
$O$ and $P$ are in $\Sigma_t$. On the other hand
$\overrightarrow{OP'}$ is in the 3-space $\Sigma_\tau$ and a pull
back from $P'$ to $S$ as renormalization is not correct. Our
practice has proven this argument.

In order to check the validity of our scheme for calculating the
relativistic LE, we are going to reexamine the core-shell system
studied by Vieira \& Letelier[6]. In Schwarzschild coordinates
$(t,r,\theta,\phi)$, the 4-metric for this system is of the form
\begin{eqnarray}
ds^2 &=& -(1-\frac{2}{r}) e^{A}dt^{2}+e^{B-A}
[(1-\frac{2}{r})^{-1}dr^{2}+r^{2}d\theta^{2}]  \nonumber \\
& &+e^{-A}r^{2}\sin^{2}\theta d\phi^{2},
\end{eqnarray}
where $A$ and $B$ are functions of $r$ and $\theta$ only. Here we
adopt nondimensional variables and take $c=1$. The metric does not
explicitly depend on $t$ and $\phi$ and has an energy constant $E$
and an angular momentum constant $L$, so test particles in free
fall are actually in a system with two degrees of freedom with an
integral $U^{\alpha}U_{\alpha}=-1$. When $A=B \equiv 0$, Eq.(7)
represents the Schwarzschild spacetime in which test particles in
free fall move in regular orbits due to integrability of the
system. For simplicity, we discuss the dynamical behavior of
geodesics in a black hole plus a dipolar shell, and let
\begin{eqnarray}
A &=& 2 \sigma \mu\upsilon, \nonumber\\
B &=& \gamma_0+ 4 \sigma \upsilon -\sigma^{2} [\mu^{2}
(1-\upsilon^{2}) +\upsilon^{2}],
\end{eqnarray}
where $\mu=r-1$ and $\upsilon=\cos \theta$. The Newtonian limit of
the model resembles the Stark problem[20,21], which is fully
integrable. However, as far as the relativistic model is
concerned, Vieira \& Letelier[6] and Saa \& Venegeroles[7] have
demonstrated strong chaos in Weyl coordinates using the
Poincar$\acute{e}$ surface of section, and Gu$\acute{e}$ron \&
Letelier[8] estimated its maximum LE
$\lambda_N=(3.2\pm0.4)\times10^{-4}$ with the classical definition
of LE (see Eq.(1)) in prolate spheroidal coordinates.
\begin{figure}[htb]
\includegraphics[scale=0.5]{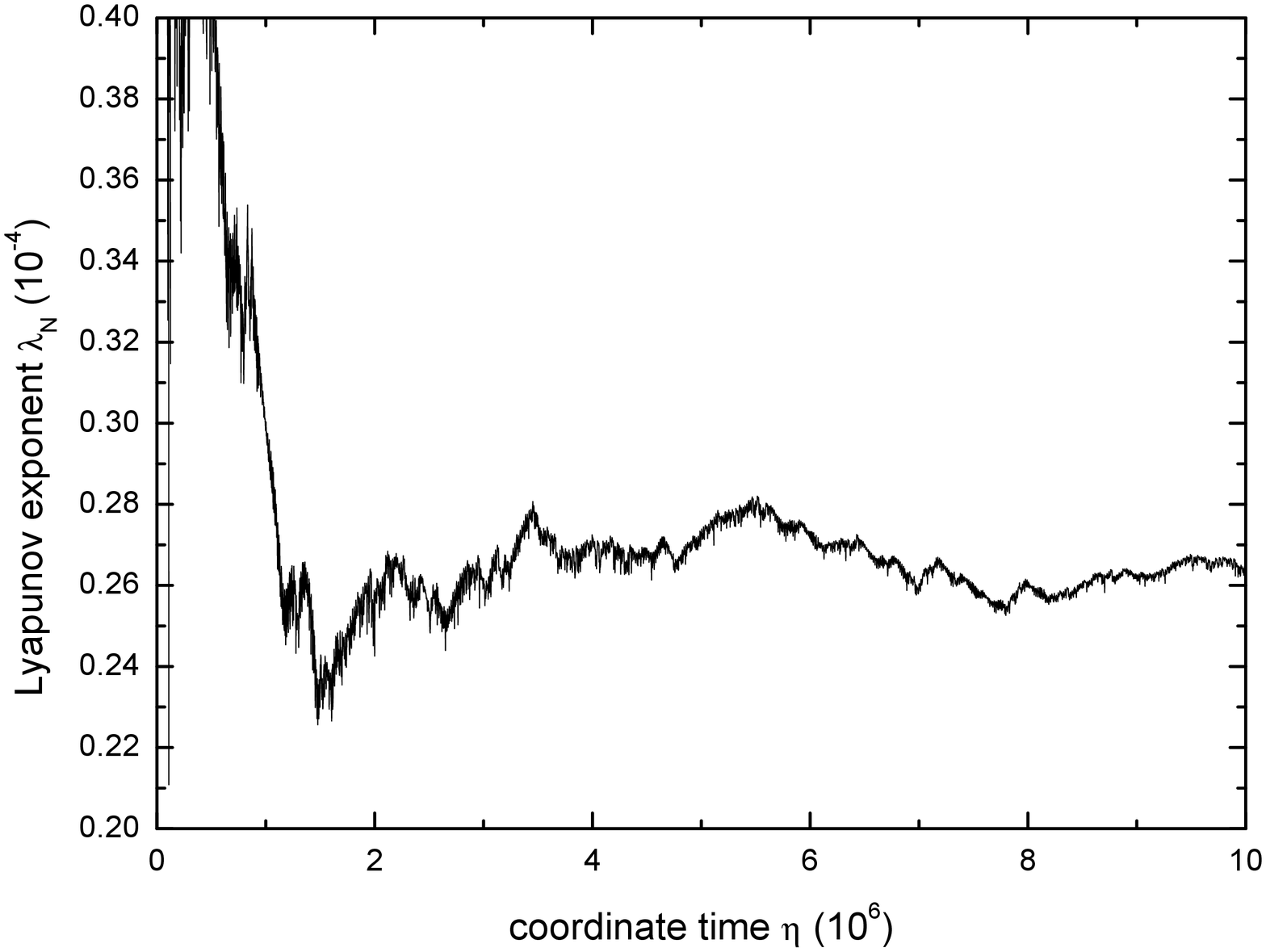}% Here is how to import EPS art
\caption{\label{fig:epsart}The maximum Newtonian Lyapunov
exponent, $\lambda_{N}$, with the time variable $\eta$. It becomes
one order of magnitude smaller after the time transformation (see
Eq.(9)).}
\end{figure}
\begin{figure}[htb]
\includegraphics[scale=0.5]{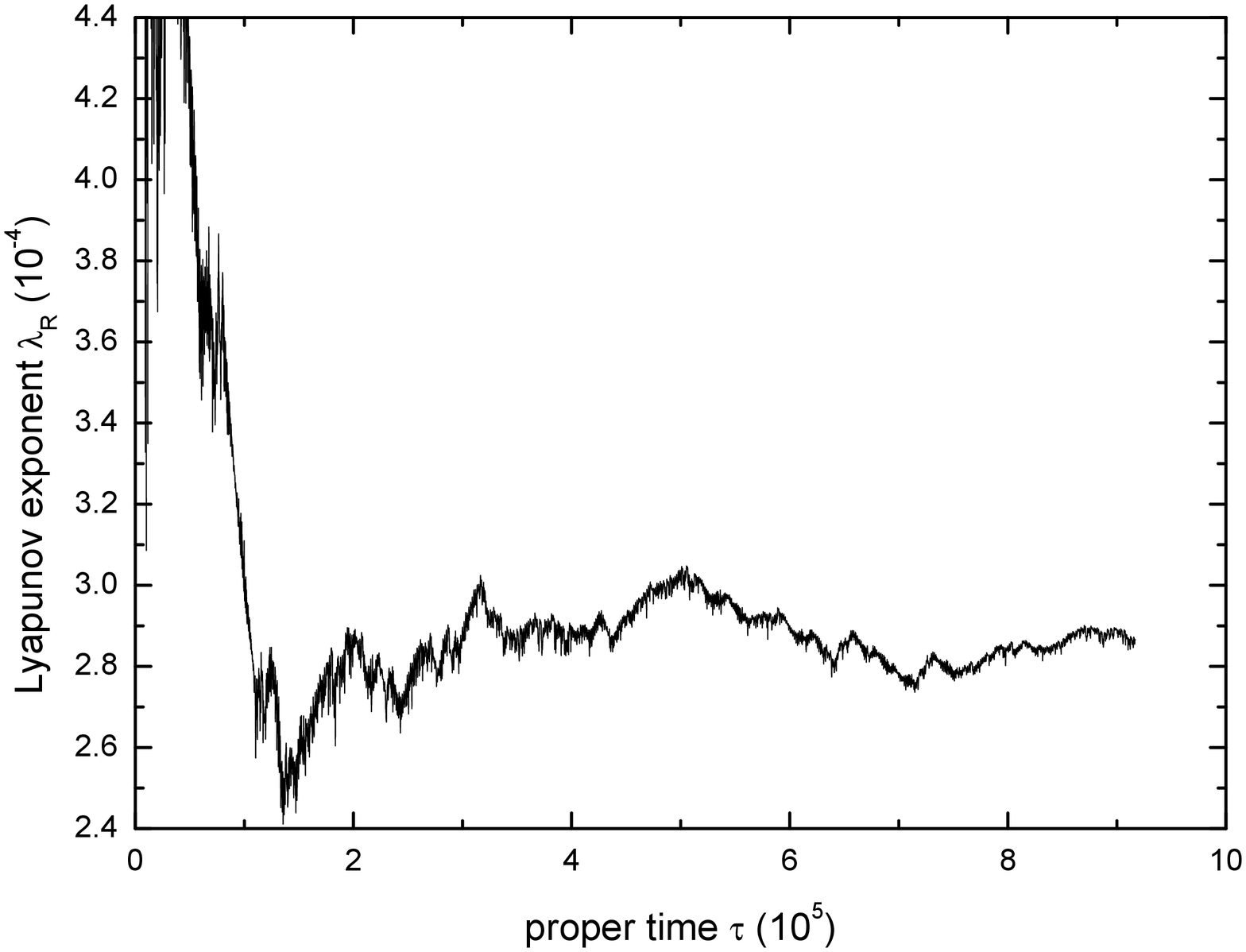}% Here is how to import EPS art
\caption{\label{fig:wide}The maximum relativistic Lyapunov
exponent, $\lambda_{R}$, with the proper time  $\tau$ of the
observer. It remains invariant after the time transformation (see
Eq.(9)). $\tau$ reaches about $9.168\times10^5$ when $\eta$ runs
$10^7$.}
\end{figure}

We choose parameters as $E=0.975$, $L=3.8$,
$\sigma=2.5\times10^{-4}$, and $\gamma_{0}=\sigma^{2}$, and the
initial conditions of the observer as $r=32$,
$\theta=\frac{\pi}{2}$, $\phi=0$, $\dot{r}=0$, and $\dot{\theta}$
from $U^{\alpha}U_{\alpha}=-1$. As to its neighbor, an initial
separation $\Delta r=-10^{-8}$ is adopted, regarded as the best
choice[3], and the others remain the same as the observer's except
$\dot{\theta}$. The values of $E$ and $L$ are carefully chosen to
assure that the trajectories of the observer and its neighbors are
bounded in a compact region. We integrate the geodesic equations
using two integrators for comparison, Runge-Kutta -Fehlberg 7(8)
and the 12th-order Adams-Cowell method, with a coordinate time
step 0.01. When $t$ reaches $10^{6}$, we find $\lambda_{N}=
\lambda'_{N}= (2.2\pm0.2)\times10^{-4}$, which is close to the
result of [8] as its integration time amounts to $10^{5}$.
Meanwhile, we get $\lambda_{R}=(2.8\pm0.2)\times10^{-4}$ by
Eq.(6). One may notice that $\lambda_{N}$ and $\lambda_{R}$ are in
the same magnitude. This is because the Euclidian distance $d'(t)$
and the invariant proper distance $\Delta L$ differ not very much
and so do the coordinate time $t$ and the proper time $\tau$. In
fact, $\tau$ runs about $9.174\times10^{5}$ when $t$ passes
through $10^{6}$. A stronger gravitation by decreasing the angular
momentum $L$ will increase the difference between $\lambda_{N}$
and $\lambda_{R}$, but the smallest stable circular orbit in
Schwarzschild spacetime is $r=6$.

To verify the invariance of $\lambda_R$, we do a time
transformation
\begin{equation}
t\rightarrow \eta=10t+r^{2}/2.
\end{equation}
We obtain $\lambda_{N}= (2.6\pm0.2) \times10^{-5}$ in the
coordinate time $\eta$, while $\lambda_{R}$ retains the original
value (see Fig.2 and Fig.3).

As a further experiment, we slightly change the model to put
$B/2\equiv A=2 \sigma \mu\upsilon$. This system is still chaotic
in Schwarzschild coordinates $(t,r,\theta,\phi)$. To remove the
singularity of this metric at the horizon, we go to Lema\^{i}tre
coordinates $(T,R,\theta,\phi)$[22] as
\begin{eqnarray}
T &=& \kappa( t+2\sqrt{2r}+
2\ln|\frac{\sqrt{r}-\sqrt{2}}{\sqrt{r}+\sqrt{2}}|), \nonumber \\
R &=& \frac{2\kappa r}{3}\sqrt{\frac{r}{2}}+T,
\end{eqnarray}
where the positive constant $\kappa$ may be chosen arbitrarily.
Our numerical tests display that the larger $\kappa$ is, the
smaller $\lambda_{N}$ becomes. Particularly, the Lema\^{i}tre
coordinates hide chaos for sufficient large $\kappa$ if
$\lambda_N$ is adopted as a chaotic index. However, our
$\lambda_{R}$ does not vary with the parameter $\kappa$.

We would like to emphasize again that the computation of
$\lambda_R$ can be applied, whether the particles move along
geodesics or not. The theory of observation in general relativity
is not relevant to the 4-acceleration of the observer. For
example, this method can be used to identify chaos in compact
binary systems[23,24].

In this letter we concentrate on calculating an invariant maximum
LE in general relativity, but this discussion can be easily
extended to find all the relativistic LEs numerically with the
technique proposed by Benettin et al.[25]  (also see [26]). They
suggest choosing an initial set of orthonormal tangent vectors as
a base of the tangent space of the observer, then compute the
evolution of the volume determined by these vectors. To avoid two
vectors getting close to each other under evolution they use
Gram-Schmidt procedure to orthonormalize the base after each time
step. The only change in a relativistic model is in computing the
evolution of a tangent vector, which can be realized by the
technique in this letter. As to the inner product in the
orthonormalizing procedure and the norm computation, one has to
use the Riemanian inner product in place of the Euclidean one.

%\begin{acknowledgments}
We are very grateful to Dr. Xin-lian Luo of Nanjing University for
his helpful discussion.  This research is supported by the Natural
Science Foundation of China under contract Nos. 10233020 and
10173007.
%\end{acknowledgments}

\end{document}